\documentclass[a4paper,11pt]{article}
\usepackage[utf8x]{inputenc}
\usepackage{authblk}
\title{Finite BRST Transformations for the Bagger-Lambert-Gustavsson Theory  }
\author[1]{ Mir Faizal{\footnote{Email: mir@maths.ox.ac.uk}}}
 \author[2]{Bhabani P  Mandal{\footnote{Email: 
bhabani.mandal@gmail.com}}}
\author[2]{ Sudhaker Upadhyay{\footnote{Email: sudhakerupadhyay@gmail.com}}}
\affil[ 1]{Mathematical Institute, University of Oxford,
Oxford, OX1 3LB, United Kingdom}
\affil[ 2]{Department of Physics,  
Banaras Hindu University,  
Varanasi-221005, India}

\begin{document}

\maketitle

\begin{abstract} 
In this paper we analyse the
 Bagger-Lambert-Gustavsson (BLG)
theory  in $\mathcal{N} =1$ superspace. Furthermore,  we will  construct the
 BRST transformations for this theory. These BRST transformations will be 
 integrated out to obtain the finite field dependent version of BRST (FFBRST) transformations. 
We will also analyse the effect of the FFBRST transformations on the  effective action.  
We will thus  show that the FFBRST transformations  can be used to relate generating functionals of the BLG theory 
in two different gauges.  
\end{abstract}
\section{Introduction}
Bagger-Lambert-Gustavsson (BLG) theory is thought to be the 
dual   boundary gauge theory to a 11-dimensional supergravity theory living on 
$AdS_4 \times S_7$. It  is thus a superconformal field theory with 
 $\mathcal{N} = 8$ supersymmetry and its  gauge group  is $SO(4)$. 
This theory has been   constructed using   
 Lie $3$-algebra \cite{1,2,3,4,5}. It has been 
analysed in $\mathcal{N} = 1$ superspace  
formalism \cite{14, ab1}.  
By complexifying the matter fields, the BLG theory can also be written as a  Chern-Simons-matter 
 theory with the gauge group $SU (2) \times SU (2)$ generated by ordinary Lie algebra. One
of the gauge groups is associated with Chern-Simons level, $k$  and the other, with $-k$ \cite{abm}. 
BLG theory only represents two M2-branes, and it has not been possible to express more than two M2-branes using 
the BLG theory. 
However,  inspired by BLG theory Aharony-Bergman-Jafferis-Maldacena (ABJM) 
theory has been constructed, and this theory represents N M2-branes \cite{abjm, 1a}. 
 The gauge symmetry in the ABJM theory 
is generated by an ordinary Lie algebra and the gauge group of this theory is $U(N)\times U(N)$.
This theory only has manifest   $\mathcal{N} = 6$ supersymmetry, which is expected to get enhanced to 
 $\mathcal{N} = 8$ supersymmetry from a variety of mechanisms \cite{abjm2}. 
In this paper we will only analyse the BLG theory using  Lie $3$-algebra, as it is more difficult to write
 the FFBRST transformations for the ABJM theory. 

It may be noted that as the BLG theory has gauge symmetry, it cannot be quantized without getting rid 
of these unphysical degrees of freedom. This 
can be done by fixing a gauge. The gauge fixing condition can be incorporated at a quantum level by 
adding  ghost and gauge fixing terms to the original classical Lagrangian. It is known that 
for  a gauge theory the new effective Lagrangian 
constructed as the sum of the original classical Lagrangian with the gauge fixing and the ghost terms, 
is invariant a new set of transformations called the BRST transformations \cite{brst, brst1}.
BRST symmetry has also been studied in non-linear gauges  \cite{nlbrst,nlbrst1}. 
In  these gauges quadratic ghost interactions are introduced and the effective theory is invariant 
under a larger algebra called the  Nakanishi-Ojima algebra \cite{no}. 
In fact, BRST symmetry for ABJM theory has also been 
studied \cite{abm}. The BRST symmetry can  be  used to project out the sub-space of 
physical 
states from the total Hilbert space. It has been 
demonstrated that the nilpotency of these 
transformations is crucial for the unitarity of the $S$-matrix in the 
$M$-theory \cite{abm}. Even though the BRST symmetry of the ABJM theory has been studied,
so far the BRST symmetry of  based on a Lie $3$-algebra has not been
studied. It would be interesting to analyse the BRST symmetry directly
based on Lie $3$-algebra because this structure has been used to study
the action for M5-branes \cite{ho}.
In this proposal  the BLG action with Nambu-Poisson 3-bracket has been
identified with  the M5-brane action with a large worldvolume three
form field. In this paper we analyse the infinitesimal BRST symmetry for the BLG theory. 

The infinitesimal global BRST transformations can be integrated out to obtain the FFBRST transformations
\cite{jm}. Various applications of these FFBRST transformations have been studied \cite{jm, ff, ff1, sud3, sd, sud4,bss, sud5}. 
  A correct prescription for the poles in the gauge field propagators in 
non-covariant gauges has been derived by connecting effective theories in covariant gauges to 
the theories in noncovariant  gauges by using FFBRST transformation \cite{sdj2}.
The divergent energy integrals in the Coulomb gauge have also been
regularized by modifying the time like propagator by using FFBRST transformation \cite{cou}.
The  Gribov-Zwanziger theory \cite{gri,zwan}, which is free from  Gribov copies and plays 
a crucial role in the non-perturbative infrared regime while it can be neglected in the perturbative 
ultraviolet  regime,  has also been related  to a theory with Gribov copies i.e. Yang-Mills theory in
Euclidean space through FFBRST transformation \cite{sud1,sud2}. 
In this paper we will also analyse the FFBRST for the BLG theory and show 
how it can be used to relate the BLG theory in two different gauges. 

\section{ BLG Theory}
In this section, first of all we review  the construction of BLG theory 
in $\mathcal{N} =1 $ superspace. 
To do that we first start from reviewing the basic properties of a Lie $3$-algebra. 
A Lie $3$-algebra ${\cal A}$ is a vector space with
 basis $T^A$, $a=1,\ldots,
{\rm dim}\,{\cal A}$, endowed with a trilinear antisymmetric
product \cite{blgblg},
\begin{equation}
  [T^A,T^B,T^C]=f^{ABC}_{D}T^D.
\end{equation}
The algebra is accompanied by an inner product,
$h^{AB}= Tr(T^A T^B)$, 
 with which indices may be raised and lowered.
The structure constants of the algebra are required to be
totally anti-symmetric, $f^{ABCD} = f^{[ABCD]}$ and satisfy the 
 the fundamental identity, 
$ f^{[ABC}{}_G f^{D]EG}{}_H = 0. $
It is also useful to define 
$ C^{AB,CD}_{EF} = f^{AB[C}_{[E} \delta^{D]}_{F]}$ \cite{blgblg}.
 These are antisymmetric in the pair of 
indices $AB$ and $CD$ and 
satisfy 
\begin{equation}
C^{AB,CD}_{EF} C^{GH,EF}_{KL} + C^{GH,AB}_{EF} C^{CD,EF}_{KL} 
+C^{CD,GH}_{EF} C^{AB,GH}_{KL} = 0.
\end{equation}

The classical Lagrangian density
 for
the BLG theory in this  superspace formalism is given by,
$ { \mathcal{L}_c} =  \mathcal{L}_{M}
 + \mathcal{L}_{CS}$,
where $\mathcal{L}_{CS}$ is   
 the Lagrangian densities for the Chern-Simons theory and 
$\mathcal{L}_{M}$ is the Lagrangian density for the matter fields.
Now, the non-Abelian    Chern-Simons   theory  
on this  superspace  can now be
 written as 
\begin{equation}
 \mathcal{L}_c =\frac{k}{4\pi} \int d^2 \,  \theta \, \,  
 Tr [ f^{ABCD}  \Gamma^{a}_{AB}    \Omega_{a CD}],
\end{equation}
where $k$ is an integer and 
\begin{eqnarray}
 \Omega_{AB a} & = & \omega_{a AB} - 
\frac{1}{6}C^{CD, EF}_{AB} \Gamma^b_{CD} \Gamma_{ab EF} \\
 \omega_{AB a} & = & \frac{1}{2} D^b D_a \Gamma_{AB b} 
- \frac{i}{2}  C^{CD, EF}_{AB}\Gamma^b_{CD}  D_b \Gamma_{a EF} 
\nonumber \\ && -
 \frac{1}{6} C^{CD, EF}_{AB} C_{EF}^{LM, NP} \Gamma^b_{CD} 
 \Gamma_{b LM } \Gamma_{a NP}, \label{omega} \\
 \Gamma_{AB ab} & = & -\frac{i}{2}  \left[D_{(a}\Gamma_{AB b)} 
- iC^{CD, EF}_{AB}\Gamma_{a CD}  \Gamma_{b EF} \right],
\end{eqnarray}
here $D_a$ is given by 
$
 D_a = \partial_a + (\gamma^\mu \partial_\mu)^b_a \theta_b
$.
The Lagrangian density for the matter fields  is given by 
\begin{eqnarray}
 \mathcal{L}_{M} &=& \frac{1}{4} \int d^2 \,  \theta \, \,  
Tr \left[ \nabla^a_{}          X^{I \dagger}          
\nabla_{a }          X_I +\mathcal{V}_{   } \right],
\end{eqnarray}
where the covariant derivatives are given by 
$
 \nabla_{a}         X^{ A I } = 
D_a  X^{ A I } + i \Gamma^{AB}_{a  }         X^{I }_B,
$
and  the potential term  given by 
$
 \mathcal{V}  = f_{ABCD}\epsilon^{IJ}\epsilon^{KL} 
 [ X^A_I  X^{B\dagger}_K  X^C_J  Y_L^{D\dagger}] 
$.
The infinitesimal gauge  transformations for these fields are written as,   
\begin{eqnarray}
\delta X^{IA }= i(\Lambda{     } X^{I })^A,  
&&  \delta  X^{IA \dagger  } 
= -i(X^{IA\dagger  }{     } \Lambda)^A, 
\nonumber \\ \delta \Gamma^A_a =(\nabla_a {     } \Lambda)^A. &&
\end{eqnarray}
The Lagrangian for 
the BLG theory is invariant under these  gauge transformations,  
$
 \delta \mathcal{L}_{BLG} = 0$, where $  \delta \mathcal{L}_{BLG} = \delta \mathcal{L}_{kcs} (\Gamma) -  
\delta \tilde{\mathcal{L}}_{-kcs} (\tilde\Gamma)  
+  \delta \mathcal{L}_M 
$.
All the degree's of freedom in the Lagrangian density for
 this    BLG  theory are not physical because it is invariant
 under gauge transformations.
So, we have to fix a gauge before doing any 
calculations. This can be done by choosing 
the following   gauge fixing conditions, $G = 0$, where  
$
G =   D^a  \Gamma_a 
$.
These  gauge fixing conditions can be
 incorporate at the quantum level by adding the following  
gauge fixing term to 
the original Lagrangian density,
\begin{equation}
\mathcal{L}_{gf} = \int d^2 \,  \theta \, \, 
 \left[f^{ABCD}b_{AB}  D^a \Gamma_{a CD} + \frac{\alpha}{2}f^{ABCD} b_{AB}  b_{CD}  
\right].
\end{equation}
The ghost term  corresponding to this gauge fixing term can be written as  
\begin{equation}
\mathcal{L}_{gh} = \int d^2 \,  \theta \, \, 
\left[ f^{ABCD}\overline{c}_{AB}   D^a \nabla_a   c_{CD}
\right].
\end{equation}

\section{BRST Symmetry }
The total Lagrangian density obtained by addition 
of the original classical Lagrangian density, the gauge 
fixing term and the ghost term can be used to construct the effective action for 
the BLG theory as, 
\begin{equation}
 S_{BLG} = \int d^3 x [ \mathcal{L}_c +\mathcal{L}_{gf}+  \mathcal{L}_{gh} ].\label{action}
\end{equation}
This effective action is used to define the  generating functional for the BLG theory as
\begin{equation}
Z =\int {\cal D}  \Gamma  {\cal D} c {\cal D} \bar c {\cal D} b \ e^{iS_{BLG}}.
\end{equation}
The total Lagrangian density given in Eq. (\ref{action}) is invariant under 
the following infinitesimal BRST transformations, 
\begin{eqnarray}
s \,\Gamma^{AB}_{a} = [\nabla_a   c]^{AB}\   , 
&&
s \,c_{AB} = - \frac{1}{2}C^{CD, EF}_{AB}{c_{CD} c_{EF}}\   ,
\nonumber \\
s \,\overline{c}^{AB} = b^{AB}\   , 
&&
s \,b^{AB} =0,  \nonumber \\ 
s \, X^{I A } = ic^{AB}  X^{I }_{B}\   , 
 &&  
s \, X^{ I A\dagger }
 =  - i  X^{I \dagger }_B c^{AB}\   ,
\end{eqnarray}
where BRST parameter $  $ is  global, infinitesimal and anticommuting in nature.

 Now, to check the nilpotency of such transformations we have  
\begin{eqnarray}
s^2 b^{AB} &=& 0,  \nonumber \\ 
 s^2 \, \overline{c}^{AB} &=& s\,  b^{AB}  = 0, \nonumber \\ 
s^2 \, c_{AB} &=& - s\, \frac{1}{2} C_{AB}^{CD, EF}  c_{CD}    c_{EF} \nonumber \\ &=& 
 \frac{1}{4} C_{AB}^{CD, EF}  C_{CD}^{LM, PT} c_{LM}    c_{PT}      c_{EF}\nonumber \\ &&  -  
\frac{1}{4}C_{AB}^{CD, EF}  C_{EF}^{LM, PT}  c_{CD}   c_{LM}    c_{PT}  = 0, \nonumber \\ 
s^2 \,\Gamma^{AB}_{a} &=&  s\, D_a c_{ AB} + s\, C^{CD,EF}_{AB}\Gamma_{CD a}    c_{ EF} \nonumber \\ 
&=& D_a [ c_{ AB} + C^{CD,EF}_{AB}\Gamma_{CD a}    c_{ EF}] \nonumber \\ && 
 + C^{CD,EF}_{AB}[D_a c_{CD} + C^{PQ,RS}_{CD}\Gamma_{PQ a}    c_{ RS}]    c_{ EF} \nonumber \\ &&
- C^{CD,EF}_{AB}\Gamma_{CD a}    D_a c_{ EF} \nonumber \\ && - C^{CD,EF}_{AB} C^{PQ,RS}_{EF}\Gamma_{CD a}    \Gamma_{a PQ }    c_{ RS} =0, \nonumber \\ 
s^2 \, X^{I  }_A  &=& is\, c_{AB}    X^{I B}\nonumber \\
 &=&  - \frac{i}{2}C^{CD, EF}_{AB}c_{CD}   c_{EF}    X^I_B \nonumber \\ && - i c_{AB}    c^{BE}    X^I_E =0,  \nonumber \\ 
s^2 \, X^{I \dagger  }_A  &=& - is\, X^{I B \dagger }    c_{AB} \nonumber \\
 &=&  \frac{i}{2}C^{CD, EF}_{AB} X^{I \dagger }_B    c_{CD}   c_{EF}  \nonumber \\ && +  i  X^{I \dagger }_E    c^{BE}   c_{AB} =0,
\end{eqnarray}
Thus, these  BRST transformations are nilpotent, 
$
s^2 =   0
$.

We can now express the  sum of the gauge fixing term and 
the ghost term as 
\begin{eqnarray}
\mathcal{L}_{gf} + \mathcal{L}_{gh} 
 &=&  \int d^2 \,  \theta \, \,  s\, \left[ f^{ABCD} \overline{c}_{AB} 
 \left(D^a  \Gamma_{aCD} 
 -  \frac{ \alpha}{2}b_{CD}\right)
\right].
\end{eqnarray}
In fact, the invariance of the 
total Lagrangian density follows from the nilpotency of the BRST transformations.
This is because $\mathcal{L}_{gh} + \mathcal{L}_{gf} $ 
can be expressed as a total BRST  variation and hence the action of $s$ on $\mathcal{L}_{gh} + \mathcal{L}_{gf} $
vanishes. 
The BRST variation of the original theory is the  gauge
 transformation with gauge fields replaced by ghosts or anti-ghosts,  respectively,
 \begin{equation}
 s \mathcal{L}_c +s \mathcal{L}_{gh} + s\mathcal{L}_{gf}  = 0.
\end{equation}
\section{FFBRST Transformation}
In this section, we construct the  FFBRST transformations for the BLG theory. In order to do that 
we first  define $\Phi^i(x, \kappa) = \Phi^{iAB}(x, \kappa) T_AT_B$, where $\Phi^{i AB}
 = (\Gamma^{AB}_a, c^{AB}, \overline{c}^{AB}, b^{AB})$, 
here all the fields  depend on  some parameter, $\kappa: 0\le \kappa \le 1$, in 
such a manner that ${\Phi^i}   (x, 0 )$ are the initial fields and
 $ {\Phi^i}   (x, 1)$ are the transformed field. 
Now, we also define $\Theta [{\Phi}   ]$ as a functional with odd Grassmann parity. This can  obtained from a
infinitesimal
field dependent parameter  through the following relation
\begin{equation}
\Theta  [{\Phi}   (x)] = \epsilon  [{\Phi}   (x)] \frac{ \exp F [{\Phi}   (x)]
-1}{F [{\Phi}   (x)]},
\end{equation} 
where 
\begin{equation}
 F = \frac{ \delta \epsilon [\Phi (x)]}{\delta
\Gamma_a (x)} s  \Gamma_a (x) +
\frac{ \delta \epsilon [\Phi (x)]}{\delta
c(x)} s  c (x) +
\frac{ \delta \epsilon [\Phi (x)]}{\delta
\overline c (x)} s  \overline c (x) +
\frac{ \delta \epsilon [\Phi (x)]}{\delta
b(x)} s b(x).
\end{equation}
Now, the  infinitesimal parameter in the BRST transformation is made field 
dependent and hence the BRST transformation can be written as
\begin{equation}
\frac{ d}{d \kappa}{\Phi^i}   (x, \kappa ) = s  {\Phi^i}    (x, \kappa )\
\epsilon [{\Phi}   (x,\kappa )],
\label{dif}
\end{equation}
where $\epsilon [{\Phi}   (x,\kappa )]$ is an infinitesimal field dependent parameter. By integrating these equations 
from $ \kappa=0$ to $\kappa=1$, it has been shown  
that the ${\Phi^i} ( x, 1) $ are related to ${\Phi^i}   (x, 0)
$ by the FFBRST transformation  as
\begin{equation}
{\Phi^i} (x, 1) = {\Phi^i}   (x, 0) + s  {\Phi^i}    (x, 0) \Theta [{\Phi}   (x)],
\end{equation}
Thus, we can write explicitly the  FFBRST transformation for the BLG theory as 
\begin{eqnarray}
f\,\Gamma^{AB}_{a} = [\nabla_a   c]^{AB}\Theta , 
&&
f\,c_{AB} = - \frac{1}{2}C^{CD, EF}_{AB}{c_{CD} c_{EF}} \Theta ,
\nonumber \\
f \,\overline{c}^{AB} = b^{AB}\Theta , 
&&
f \,b^{AB} =0, 
 \nonumber \\ 
f \, X^{I A } = ic^{AB}  X^{I }_{B}\Theta , 
 &&  
f \, X^{ I A\dagger }
 =  - i  X^{I \dagger }_B c^{AB}\Theta.
\end{eqnarray}
The FFBRST transformation is symmetry of the action $S_{BLG}$ only but not of the generating functional 
 as the Jacobian for path integral measure  in the expression of generating functional  is not invariant
under it. 
Under FFBRST transformation
Jacobian changes as  
$
{\cal D}{\Phi^i}    =J[{\Phi}   (\kappa)] {\cal D}{\Phi^i}   (\kappa). \label{jac}
$            
It has been shown  that this nontrivial Jacobian can be replaced within the 
functional integral as
\begin{equation}
J[{\Phi}   (\kappa)] \rightarrow e^{iS_1[{\Phi}   (\kappa)]},
\end{equation}
where $S_1[{\Phi}   (\kappa)]$ is some local functional of ${\Phi^i}$. The condition
 for existence of $S_1$ is
\begin{eqnarray}
\int d^3x d^2 \theta \,\, \left[\frac{1}{J (\kappa )}\frac{d J (\kappa )}{d\kappa} -i\frac{dS_1}{d\kappa}\right] =0.
\label{mcond}
\end{eqnarray}
To calculate the 
infinitesimal change in Jacobian we 
use the following expression, 
 \begin{eqnarray} 
 \frac{1}{J(\kappa)}\frac{dJ(\kappa)}{d\kappa}&=& 
 -\int d^3x d^2 \theta \,\,      \left[  s  \Gamma_a (x) \frac{ \delta \epsilon [\Phi (x, k)]}{\delta
\Gamma_a (x, k)} -s  c (x, k)
\frac{ \delta \epsilon [\Phi (x)]}{\delta
c(x, k)} \right. \nonumber \\ &&\left.   - s  \overline c (x, k)
\frac{ \delta \epsilon [\Phi (x, k)]}{\delta
\overline c (x, k)}  +s b(x, k)
\frac{ \delta \epsilon [\Phi (x, k)]}{\delta
b(x, k)} \right].\label{jaceva}
\end{eqnarray}
\section{ Relating  Different Gauges} 
In this section, we show explicitly that how FFBRST transformation can be used to analyse the
BLG theory in two different gauges. It is possible to take different gauges for 
the BLG theory. For example, we can take a non-linear gauge in the BLG theory, 
which is similar to a non-linear gauge in Yang-Mills theories. The sum of the gauge fixing and ghost terms 
for this non-linear gauge can be written as 
\begin{eqnarray}
 \mathcal{L}_{gh} + \mathcal{L}_{gf}  &=&\int d^2 \,  \theta \, \, 
f^{ABMN} \left[ [b_{AB}  D^a \Gamma_{a MN} \right. \nonumber \\ && \left.   + \frac{\alpha}{2}b_{AB}b_{MN}
+ \frac{1}{2}\overline{c}_{AB}   D^a \nabla_{a}   c_{MN} \right. \nonumber \\ && \left. 
+ \frac{1}{8} C^{CD, EF}_{AB} C^{IJ, KL}_{MN}{\overline c_{CD} c_{EF}}{\overline c_{IJ} c_{KL}}
\right].
\end{eqnarray}
The non-linear BRST transformation are now given by 
\begin{eqnarray}
s \,\overline{c}_{AB} &=& b_{AB}\   - \frac{1}{2}C^{CD, EF}_{AB}{\overline c_{CD} c_{EF}}\   , 
 \nonumber \\ 
s \,\Gamma^{AB}_{a}&=&[\nabla_a   c]^{AB}\   , 
 \nonumber \\ 
s \,b_{AB}&=&  - \frac{1}{2}C^{CD, EF}_{AB}{c_{CD} b_{EF}} \    
-  \frac{1}{8}C^{CD, EF}_{AB} C_{EF}^{LM, NP} c_{CD}
 c_{ LM } \overline c_{ NP}\   ,
 \nonumber \\ 
s \,c_{AB}&=&  - \frac{1}{2}C^{CD, EF}_{AB}{c_{CD} c_{EF}} \   ,
 \nonumber \\ 
s \, X^{I A } &=& ic^{AB}  X^{I }_{B}\   ,
\nonumber \\ 
s \, X^{ I A\dagger }
&=&  - i  X^{I \dagger }_B c^{AB}\   .
\end{eqnarray}
Just like in the linear case here again we can show that these transformations are nilpotent, $s^2 = 0$.  
The non-linear finite BRST transformations are given by 
\begin{eqnarray}
f \,\overline{c}_{AB} &=& b_{AB}\Theta - \frac{1}{2}C^{CD, EF}_{AB}{\overline c_{CD} c_{EF}}\Theta , 
 \nonumber \\ 
f \,\Gamma^{AB}_{a}&=&[\nabla_a   c]^{AB}\Theta , 
 \nonumber \\ 
f \,b_{AB}&=&  - \frac{1}{2}C^{CD, EF}_{AB}{c_{CD} b_{EF}}\Theta   
-  \frac{1}{8}C^{CD, EF}_{AB} C_{EF}^{LM, NP} c_{CD}  
 c_{ LM } \overline c_{ NP}\Theta ,
 \nonumber \\ 
f \,c_{AB}&=&  - \frac{1}{2}C^{CD, EF}_{AB}{c_{CD} c_{EF}} \Theta ,
 \nonumber \\ 
f \, X^{I A } &=& ic^{AB}  X^{I }_{B}\Theta,
\nonumber \\ 
f \, X^{ I A\dagger }
&=&  - i  X^{I \dagger }_B c^{AB}\Theta .
\end{eqnarray}
 Let the linear and the non-linear  gauges be represented by 
 $G^{AB}_1[\Gamma]$ and $G^{AB}_2[\Gamma]$   and let $(sG_1)^{AB}$ and $(sG_2)^{AB}$ be the linear BRST and 
non-linear BRST transformations of these gauge fixing conditions, respectively. 
The infinitesimal field dependent  BRST parameter is now chosen to be 
\begin{eqnarray}
\epsilon [\Phi] = 
i\gamma\int d^3x  d^2 \theta \, \, \left[ f^{ABCD}\bar c_{AB} \left( G_{CD 1} - G_{CD 2}\right)\right], 
\end{eqnarray}
where $\gamma$ is an arbitrary constant parameter.
Using expression for the change in Jacobian for this $\epsilon [\Phi]$ is calculated as
\begin{eqnarray}
\frac{1}{J}\frac{dJ}{d\kappa} &=&i\gamma\int d^3x d^2 \theta \,\,  f^{ABCD}\left[ b_{AB}G_{CD 1} - b_{AB}G_{CD 2}
\right.
\nonumber\\
 &-&\left.( sG_{CD 1} - sG_{CD 2})\bar c_{AB}\right],\nonumber\\
&=&i\gamma\int d^3x d^2 \theta \,\,  f^{ABCD}\left[ b_{AB}G_{CD 1} - b_{AB}G_{CD 2}
\right.
\nonumber\\
 &+&\left.\bar c_{AB}( sG_{CD 1} - sG_{CD 2})\right].
\end{eqnarray}
Now, an ansatz for $S_1$, 
\begin{eqnarray}
S_1 &=&\int d^3 x d^2 \theta \,\, f^{ABCD}[\xi_1 (\kappa) b_{AB}G_{CD 1}+  \xi_2 (\kappa) b_{AB}G_{CD 2} \nonumber \\ && +\xi_3(\kappa) \overline c_{AB}
s G_{CD 1} +\xi_4(\kappa)\overline c_{AB } sG_{CD 2} ],
\end{eqnarray}
where arbitrary parameters ($\xi_i(i=1,2,3,4)$) and all fields ($ b^a, \Gamma^a, c^a, \bar c^a$) depend on $\kappa$.
The parameters $\xi_i (\kappa)$ also satisfy the following initial boundary condition
\begin{equation}
\xi_i (\kappa =0)=0.\label{boun}
\end{equation}
Therefore, using Eq. (\ref{dif}) the differentiation of the above equation w. r. to $\kappa$  gives 
\begin{eqnarray}
\frac{dS_1}{d\kappa}&=&\int  d^3x d^2 \theta \,\, f^{ABCD}[\xi'_1  b_{AB}G_{CD 1} 
+\xi_1 b_{AB}
s G_{CD 1}\epsilon +  \xi'_2  b_{AB}G_{CD 2}\nonumber \\ 
 &+&\xi_2 b_{AB}
s G_{CD 2} \epsilon +\xi'_3 
\overline c_{AB}
s G_{CD 1} -\xi_3 b_{AB}
s G_{CD 1}\epsilon  \nonumber\\
&+&\xi'_4 \overline c_{AB} sG_{CD 2}
  -\xi_4 b_{AB}
s G_{CD 2} \epsilon],\nonumber\\
 &=&\int  d^3x d^2 \theta \,\, f^{ABCD}[\xi'_1  b_{AB}G_{CD 1} +  \xi'_2  b_{AB}G_{CS 2}\nonumber \\ & 
 +&\xi'_3 
\overline c_{AB}
s G_{CD 1} +\xi'_4 \overline c_{AB} sG_{CD 2} \nonumber\\
&+
&(\xi_1 -\xi_3 )b_{AB}
s G_{CD 1}\epsilon +(\xi_2 -\xi_4) b_{AB}
s G_{CD 2} \epsilon].
\end{eqnarray}
To write the Jacobian $ J\rightarrow e^{iS_1}$ the following condition (as mentioned in Eq. (\ref{mcond})) is to be satisfied,
\begin{eqnarray}
&&\int d^3x d^2 \theta \,\, \left[ 
f^{ABCD}[(\xi'_1-\gamma)  b_{AB}G_{CD 1}+( \xi'_2  +\gamma ) b_{AB}G_{CD 2}
\right.\nonumber\\
&&\left. + ( \xi'_3 -\gamma )
\overline c_{AB } sG_{CD 1}  + (\xi'_4 +\gamma ) \overline c_{AB} sG_{CD 2} 
\right.\nonumber\\
&&\left.+(\xi_1 -\xi_3 )b_{AB }sG_{CD 1}\epsilon  +(\xi_2 -\xi_4) b_{AB }sG_{CD 2}\epsilon  \right]=0.
\end{eqnarray}
Thus, we get 
$
 \xi'_1 -\gamma =0,  \, \xi_2' +\gamma =0,\, \xi'_3 -\gamma =0,\, \xi'_4 +\gamma =0, \, \xi_1 -\xi_3 =  0,\, \xi_2 -\xi_4 =0,
$ on equating the coeffiecients of the above expression.
The solutions of above equations satisfying initial condition given in Eq. (\ref{boun}) for $\gamma =1$ are 
\begin{equation}
\xi_1 = \kappa, \, \xi_2 = -\kappa, \, \xi_3 = \kappa, \, \xi_4 = -\kappa.
\end{equation}
Now, by adding $S_1 (\kappa =1 )$ to the original action having 
gauge condition $G_{CD 2}$ and corresponding ghost term, we get the finial action 
in other gauge $G_{CD 1}$ as $S_{f} = S_{BLG } + S_1$.
Thus we see that,
$Z$ transforms under FFBRST transformations to 
\begin{equation}
 Z_f = \int {\cal D} \Gamma {\cal D} c {\cal D} \bar c {\cal D} b \ e^{iS_f},
\end{equation}
which is nothing but the generating functional for BLG theory in other gauge $G_{CD 1}$.
This connection is also true for reverse manner i.e. when one starts the theory  with gauge condition $G_{CD 1}$
and then FFBRST formulation changes the theory in the gauge condition $G_{CD 2}$ with appropriate ghost terms.

Thus, the FFBRST formulation can be used to analyse the BLG theory in two different gauges. 

\section{Conclusion}
In this paper we have analysed the BLG theory in $\mathcal{N} =1$ 
superspace formalism. As theory had gauge degrees of freedom, we 
fixed a gauge to quantize it. This gauge fixing condition was incorporated at a quantum level by 
adding gauge fixing and ghost terms to the original classical Lagrangian. 
The new effective Lagrangian thus obtained was invariant under a new set of transformations 
called the BRST transformations. We explicitly wrote the BRST transformations for the BLG theory. 
These 
 BRST transformations were  integrated out to construct the FFBRST
transformations. These transformations were constructed by 
constructing a functional with odd Grassmann parity on the
gauge, ghosts, anti-ghosts and auxiliary fields.  It 
 did not depend on spacetime explicitly. FFBRST
transformations were  also found to be the symmetry of the effective  action. 
However, FFBRST transformations did not leave the path integral measure 
invariant thus were shown to connect  the generating functionals
of two different effective field theories with suitable choice of the finite field dependent
parameter.

It is known that for gauge theories in non-linear gauge the effective Lagrangian is invariant under a larger algebra 
called 
Nakanishi-Ojima algebra \cite{no}. It is also known that this algebra is broken by ghost condensation \cite{gc, gca}. 
It will be interesting to analyse the existence of Nakanishi-Ojima algebra and its subsequent 
breaking in the BLG theory. It may be noted that even for regular Yang-Mills theories the FFBRST for  non-linear 
gauges have not been analysed before. It will be interesting to analyse the FFBRST transformation for both 
regular Yang-Mills theory and the BLG theory, when the Nakanishi-Ojima algebra is broken by ghost condensation. 
\section*{Acknowledgments}
One of us (SU) gratefully acknowledges the financial support from the Council of Scientific and Industrial Research
(CSIR), India, under the SRF scheme.

\end{document}